\documentclass[preprint,superscriptaddress]{revtex4}

\usepackage{amsmath}
\usepackage{amssymb}
\usepackage{epsfig}
\usepackage{graphicx}
\usepackage{wasysym}
\usepackage{bbm}
\usepackage{color}
\usepackage{color}

\renewcommand{\v}[1]{{\bf #1}}

\newcommand{\w}{{\omega}}

\def\eqa{\begin{eqnarray}}
\def\eea{\end{eqnarray}}
\newcommand{\eq}{\begin{equation}}
\newcommand{\ee}{\end{equation}}
\newcommand{\nn}{\nonumber\\}

\newcommand{\<}{\langle}
\renewcommand{\>}{\rangle}
\newcommand{\Tr}{{\rm Tr}}

\renewcommand{\Im}{{\rm Im}}

\newcommand{\p}{\partial}

\newcommand{\ua}{\uparrow}
\newcommand{\da}{\downarrow}
\newcommand{\ra}{\rightarrow}

\newcommand{\al}{\alpha}
\newcommand{\bt}{\beta}
\newcommand{\del}{\delta}

\newcommand{\ga}{\gamma}
\newcommand{\Ga}{\Gamma}

\newcommand{\la}{\lambda}
\newcommand{\La}{\Lambda}

\newcommand{\si}{\sigma}

\begin{document}

\title{High $T_c$ superconductivity at the FeSe/SrTiO$_3$ Interface}
\author{Fa Wang}
\affiliation{Department of Physics, Massachusetts Institute of Technology, Cambridge, Massachusetts 02139, USA}
\author{Yuan-Yuan Xiang}
\affiliation{National Lab of Solid State Microstructures, Nanjing University, Nanjing, 210093, China}
\author{Da Wang}
\affiliation{National Lab of Solid State Microstructures, Nanjing
University, Nanjing, 210093, China}
\author{Qiang-Hua Wang}
\affiliation{National Lab of Solid State Microstructures, Nanjing University, Nanjing, 210093, China}
\author{Dung-Hai Lee}
\affiliation{Department of Physics, University of California at Berkeley, Berkeley, California 94720, USA}
\affiliation{Materials Sciences Division, Lawrence Berkeley National Laboratory, Berkeley, California 94720, USA}

\date{\today}

\begin{abstract}
In a recent experiment the superconducting gap of a single unit
cell thick FeSe film on SrTiO$_3$ substrate is observed by
scanning tunneling spectroscopy and angle-resolved photoemission
spectroscopy. The value of the superconducting gap is much larger
than that of the bulk FeSe under ambient pressure. In this paper
we study the effects of screening due to the ferroelectric phonons
on Cooper pairing. We conclude it can significantly enhance the
energy scale of Cooper pairing and even change the pairing
symmetry. Our results also raise some concerns on whether phonons
can be completely ignored for bulk iron-based superconductors.

\end{abstract}
\maketitle

\section{Introduction}
In a recent experiment\cite{xue} a single unit cell thick FeSe
film is grown on the TiO$_2$ terminated (001) surface of
SrTiO$_3$(STO) by molecular beam epitaxy. Two gaps ($\sim 10 $ meV
and $20$ meV) are observed by scanning tunneling microscopy (STM)
at low temperatures. At present there is no transport data and
$T_c$ is not determined. Subsequently an angle-resolved
photoemission spectroscopy (ARPES) result appears.\cite{xj}
According to this report, there is only electron Fermi surface,
suggesting FeSe is electron doped. The observed electron pockets
are nearly circular, and an approximately constant ($\sim 15$ meV)
superconducting gap is detected on them.  Significantly, by
studying the temperature dependence of the energy gap, an estimate
of $T_c$, namely, $55\pm 5$ K, is obtained.  Currently there is no
explanation of the two gap versus one gap discrepancy between the
two measurements.

There are many unanswered questions. For examples, (1) What caused
the FeSe doping? (2) Can the discrepancy between Ref.\cite{xue}
and Ref.\cite{xj} be due to surface doping caused by the sample
treatment prior to the ARPES measurement? (3) Are there buried,
hence not yet detected, interface metallic bands?\cite{xj} (4) Are
there interface ferroelectric ordering? (5) How strong is the
coupling between the electrons in FeSe and the ferroelectric (FE)
phonons in STO? Due to the uncertainty of the surface doping we
shall study both undoped and electron doped FeSe.


\begin{figure}
\includegraphics[scale=.3]{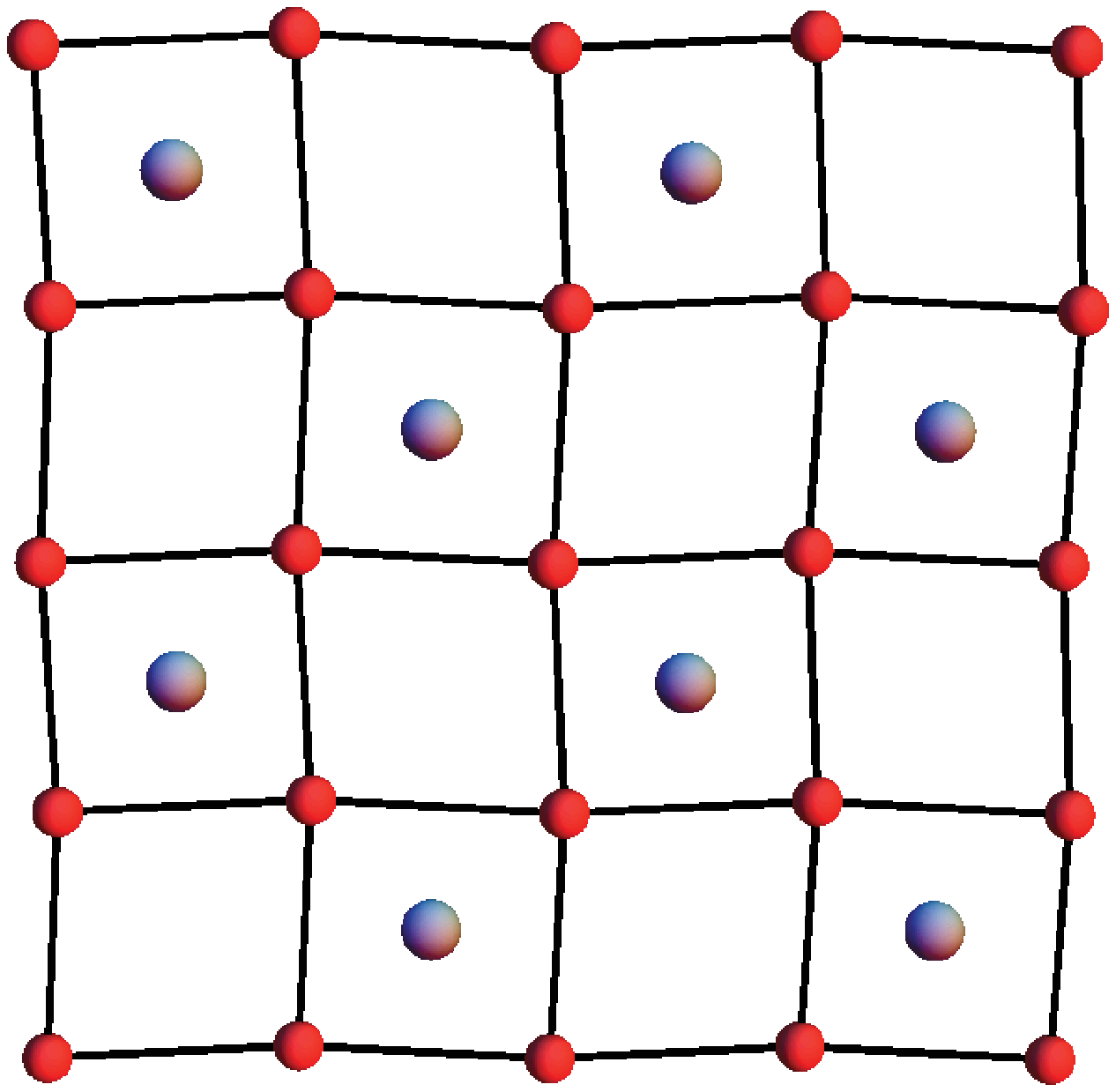}\hspace{-1in}\includegraphics[scale=.4]{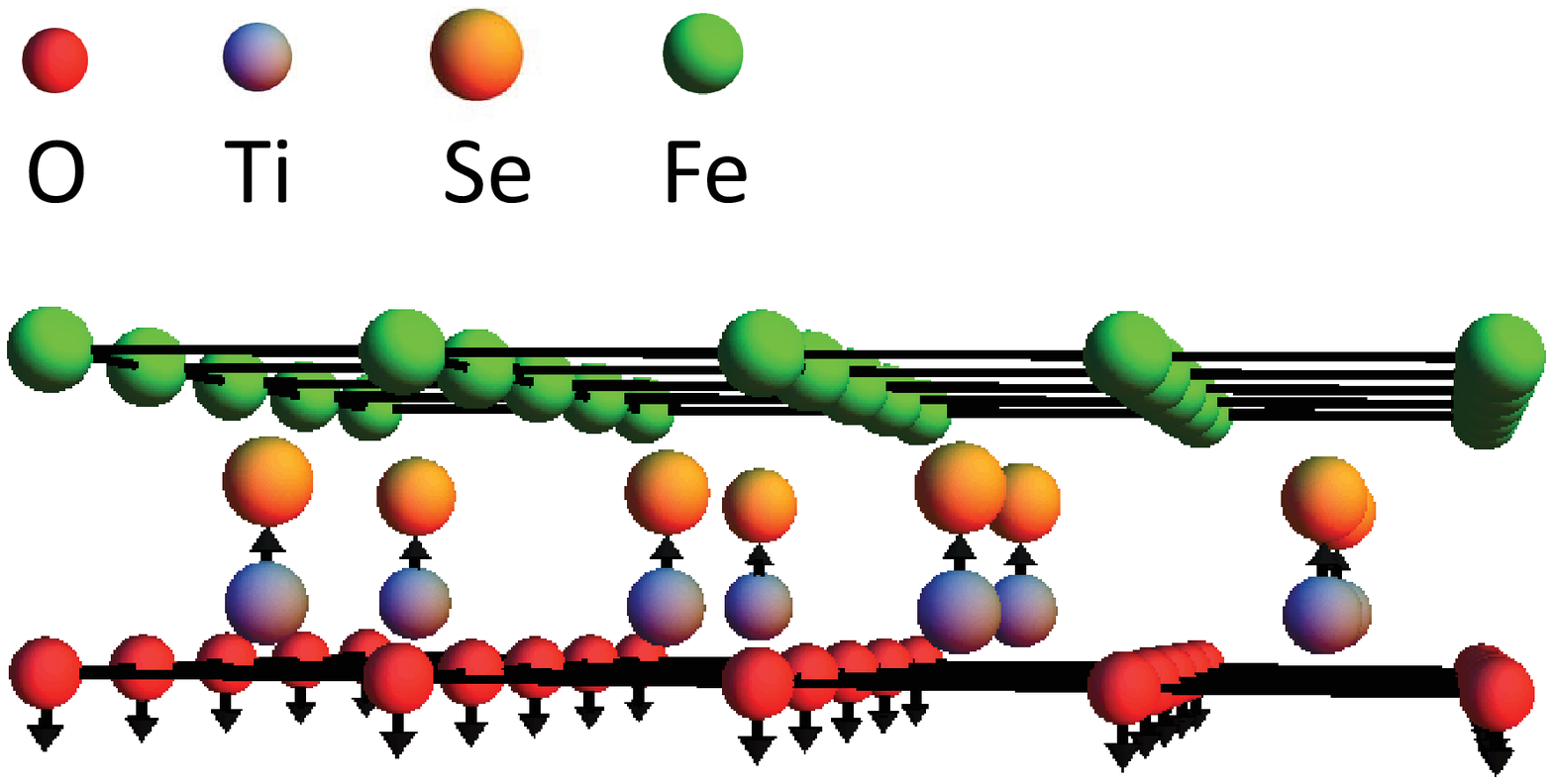}
\caption{Left panel: an example of AFD distortion in the TiO$_2$
plane. Right panel: a caricature of the frozen FE phonon with the
atomic displacement in the (001) direction. \label{phonon}}
\end{figure}

STO is a ``quantum paraelectric'' insulator. The huge dielectric
constant at low temperatures is due to the ionic movements. The associated phonon,
the FE phonon, 
involve the relative displacement of the Ti and O atoms. An example of such displacement is shown in
the right panel of Fig.~\ref{phonon}. This phonon mode is soft at the zone
center.  
At low temperatures bulk STO also undergoes an antiferrodistortive (AFD) transition.
This distortion involves alternating clockwise and counterclockwise rotation of oxygens about
titanium (see the left panel of Fig.~\ref{phonon})
In a recent local density functional study\cite{lu} it is found,
aside from zone folding, even static AFD distortion has little
effect on the bandstructure. Therefore we shall ignore the AFD
phonon in the following discussions.

In the rest of the paper we perform a two-stage renormalization group (RG) study.\cite{Kivelson,Fu}
The first stage is a functional renormalization group (FRG)\cite{frg1,frg} calculation.
It determines the most important electronic scattering processes at low energies. As found in
previous studies\cite{frg,Chubukov,wl} for energies lower than the magnetic fluctuation energy scale, $\Lambda_e$,
the strongest electron-electron scattering is in the Cooper channel (for superconducting samples). This is
used as the input for the the second stage analysis. Here we study the effect of FE phonons at energies
lower than the maximum phonon frequency $\Lambda_{ph}<\Lambda_e$. This is done by solving a generalized Eliashberg
equation. The FE phonons have two important effects on the FeSe electrons. The first is to mix states separated
by momentum $(\pi,\pi)$ in the unfolded Brillouin zone. This is because a frozen FE phonon breaks
the $z\leftrightarrow -z$ reflection symmetry, the symmetry that enables the Brillouin zone unfolding.
In addition, the soft zone center FE phonons can screen the intra-pocket electron-electron repulsion.
\section{The main results}
To present the underlying physics, let's  simplify the electronic structure by retaining only two Fermi pockets: the hole and electron pockets for the undoped FeSe/STO, and the two electron pockets for the doped FeSe/STO. The two main effects of FE phonon discussed earlier is described by an intra-pocket and an inter-pocket coupling constant, $\la_{\rm intra}$ and $\la_{\rm inter}$. 
The  superconducting transition temperature determined from the  two-stage RG approach is given as follows.
\eqa &&{\rm Odd~sign~pairing:}~~~T_c=\Lambda_{\rm ph}\cdot \exp\left\{ -
\frac{1+\lambda_+}{\rm{Max}\left[\lambda_{-} - V_{-}^*\left(1+\lambda_{-}
\frac{\langle\omega\rangle}{\Lambda_{\rm ph}}\right),0\right]} \right\}\nn
&&{\rm Even~sign~pairing:}~~~T_c=\Lambda_{\rm ph}\cdot \exp\left\{ -
\frac{1+\lambda_+}{\rm{Max}\left[\lambda_+ - V_+^*\left(1+
\frac{\langle\omega\rangle}{\Lambda_{\rm ph}}\right),0\right]} \right\}
\label{equ:tc} \eea where \eqa &&\lambda_\pm=\lambda_{\rm
intra}\pm\lambda_{\rm inter}\nn &&V_\pm=V_{\rm intra}\pm V_{\rm
inter}\nn &&V_\pm^*=V_\pm/[1+V_\pm\ln(\Lambda_e/\Lambda_{\rm
ph})].\eea
Here ``Odd''/``Even'' sign pairing means the gap function has opposite/same sign on the two Fermi pockets. $V_{\rm intra}$ and $V_{\rm inter}$ are the average of the intra and inter-pocket Cooper scattering strength obtained from the first stage calculation. $\lambda_{intra}$ and $\lambda_{inter}$ are the strength of the phonon mediated Cooper scattering entering the second stage calculation. $\langle\omega\rangle$ is a weighted average\cite{McMillan68} of the phonon frequency satisfying  $0<\langle\omega\rangle/\Lambda_{\rm ph}<1$.
In obtaining the above result we have assumed that $1+V_\pm\ln(\Lambda_e/\Lambda_{\rm
ph})>0$, i.e., the pure electronic driven $T_c$ os lower than $\Lambda_{\rm ph}/k_B$.

According to Eq.(\ref{equ:tc}) increases $\lambda_{\rm
intra}-\lambda_{\rm inter}$, and $V_{\rm inter}-V_{\rm intra}$
raises the $T_c$ for the odd sign  pairing. On the other hand, to
raise the $T_c$ of the even sign pairing we need to increase
$\lambda_{\rm intra}+\lambda_{\rm inter}$ but decrease  $V_{\rm
intra}+V_{\rm inter}$, i.e., \eqa &&\left(\lambda_{\rm
intra}-\lambda_{\rm inter}\right)\nearrow~{\rm~and~}\left(V_{\rm
inter}-V_{\rm intra}\right)\nearrow~\Rightarrow T_c^{\rm
odd}\nearrow\nn &&\left(\lambda_{\rm intra}+\lambda_{\rm
inter}\right)\nearrow~{\rm~and~}\left(V_{\rm inter}+V_{\rm
intra}\right)\searrow~\Rightarrow T_c^{\rm even}\nearrow.
\label{mtc}\eea The physics behind Eq.(\ref{mtc}) is rather
simple. Odd/Even sign pairing {{\em requires}} the inter-pocket
Cooper scattering to be repulsive/attractive. Since phonon
mediated scattering is necessarily attractive, it follows that
strong inter-pocket electron-phonon interaction enhances even sign
pairing while suppress the odd sign pairing. Because attractive
intra-pocket scattering strengthen both even and odd sign pairing,
strong intra-pocket electron phonon interaction is beneficial to
both.

\subsection{The undoped FeSe/STO}
\begin{figure}
\includegraphics[scale=.56]{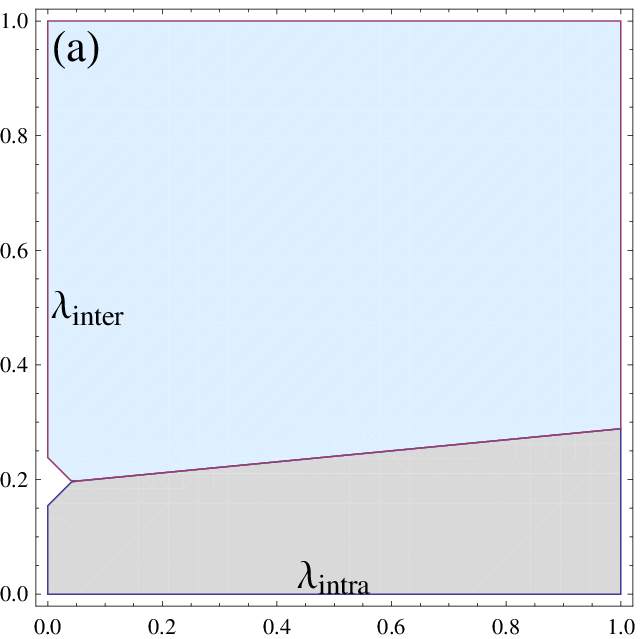}\hspace{1.in}\includegraphics[scale=.56]{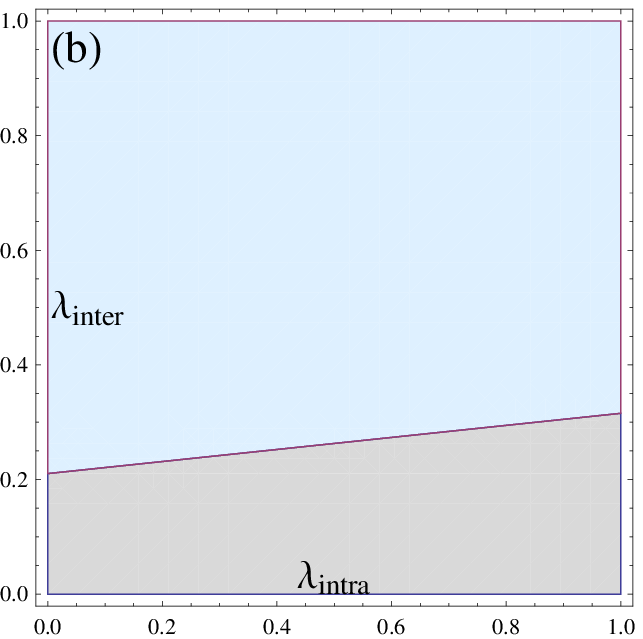}
\includegraphics[scale=.275]{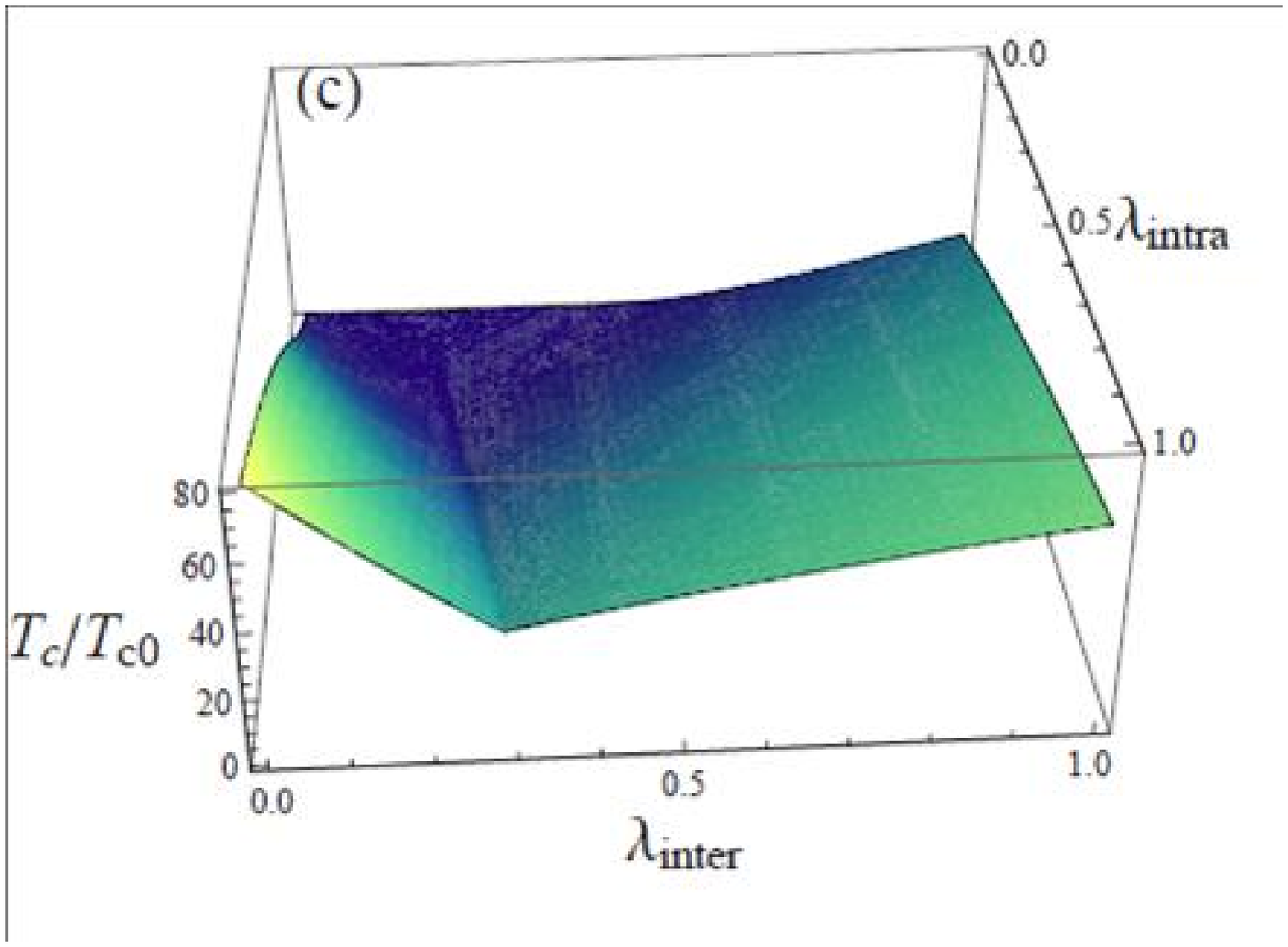}\includegraphics[scale=.275]{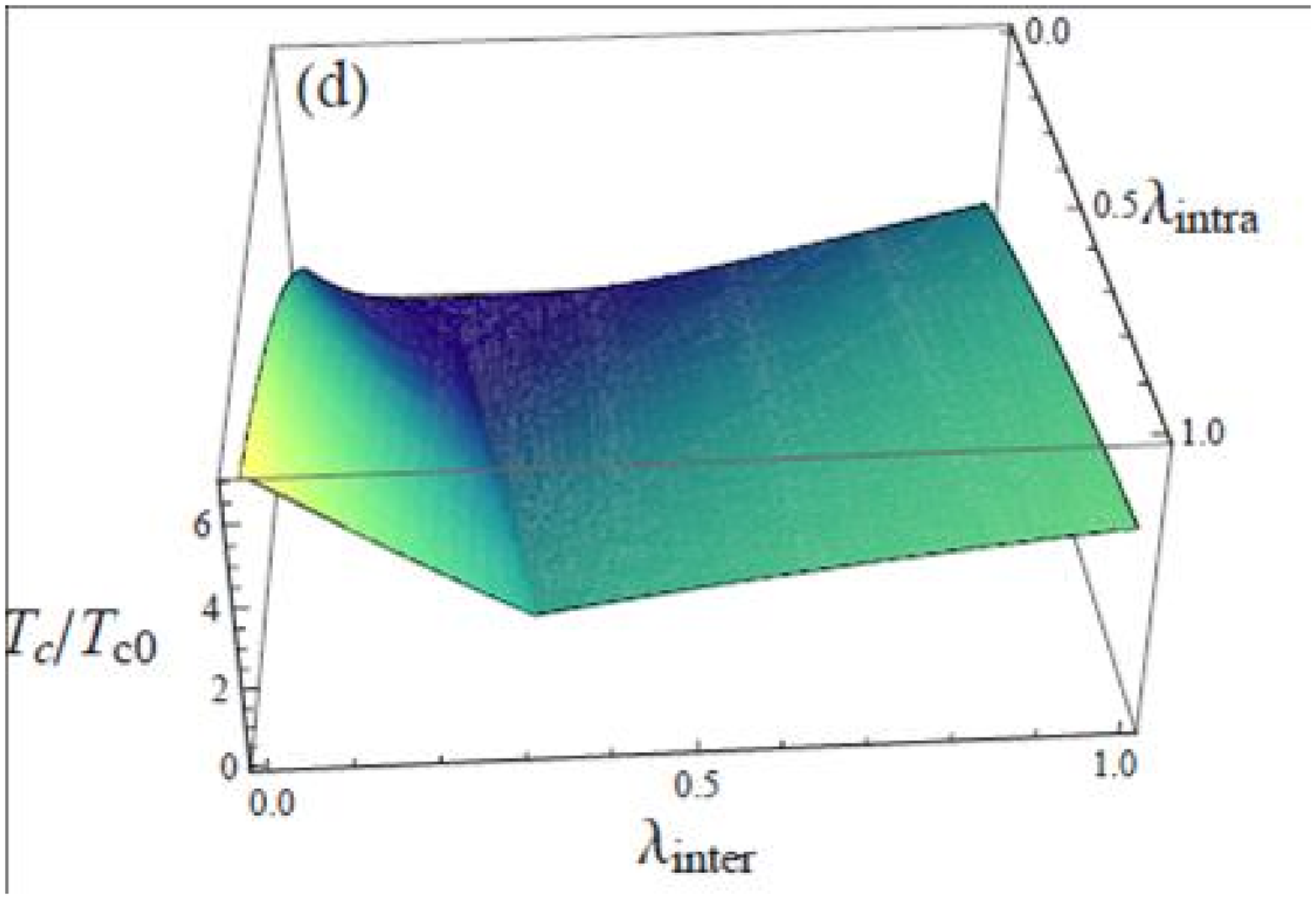}
\caption{(a,b) the phase diagram. Light blue/gray denote even/odd
sign pairing, respectively. (c,d) the $T_c$ enhancement factor
$T_c/T_{c0}$ where $T_{c0}$ is the superconducting transition
temperature in the absence of the electron-phonon interaction. The
parameters we used to construct the figures are
$\Lambda_e/\Lambda_{ph}=2$, $\langle\w\rangle/\Lambda_{\rm
ph}=0.5$, and $V_{\rm inter}=0.2$, $V_{\rm intra}=0.05$  for (a)
and (c), and  $V_{\rm inter}=0.2$, $V_{\rm intra}=-0.05$, for (b)
and (d). The small triangular region near the lower left hand
corner of  panel (a) is non-superconducting.\label{eleph}}
\end{figure}

For undoped FeSe/STO the first stage RG generates $V_{\rm
inter}>0$ hence favoring odd sign ($S_{+-}$ )
pairing.\cite{Mazin,frg,Chubukov} This is caused by the
antiferromagnetic fluctuation. As discussed above, $\la_{\rm
intra}$ enhances the $S_{+-}$ pairing while $\la_{\rm inter}$
weakens it. Setting $V_{\rm inter}=0.2$ and $V_{\rm intra}=\pm
0.05$, the ``phase diagrams'' as a function of $\lambda_{\rm
intra}$ and $\lambda_{\rm inter}$ is shown in
Fig.~\ref{eleph}(a,b). The associated $T_c$ enhancement factor is
shown in Fig.~\ref{eleph}(c,d). Note the magnitude of the $T_c$
enhancement differs by approximately an order of magnitude in
Fig.~\ref{eleph}(c,d) by merely reversing the sign of $V_{\rm
intra}$. Clearly this quantity is not something we can confidently
predict. What is robust is the fact that when $\lambda_{\rm
inter}>>\lambda_{\rm intra}$ electron-phonon interaction
stabilizes even sign pairing. Conversely for $\lambda_{\rm
intra}>>\lambda_{\rm inter}$ odd sign pairing is favored. Since
for undoped FeSe/STO the FE phonons mainly cause intra-pocket
electron scattering ( $\lambda_{\rm intra}>>\lambda_{\rm inter}$),
we expect the  electron-phonon interaction to strengthen the odd
sign, in this case $S_{+-}$, pairing.
\subsection{The electron doped FeSe/STO}
\begin{figure}
\includegraphics[scale=1.2]{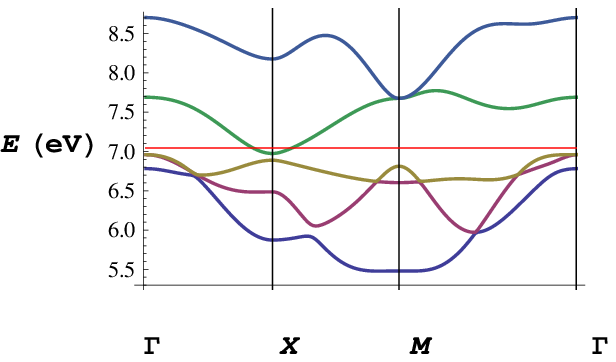}\hspace{.5in}\includegraphics[scale=.5]{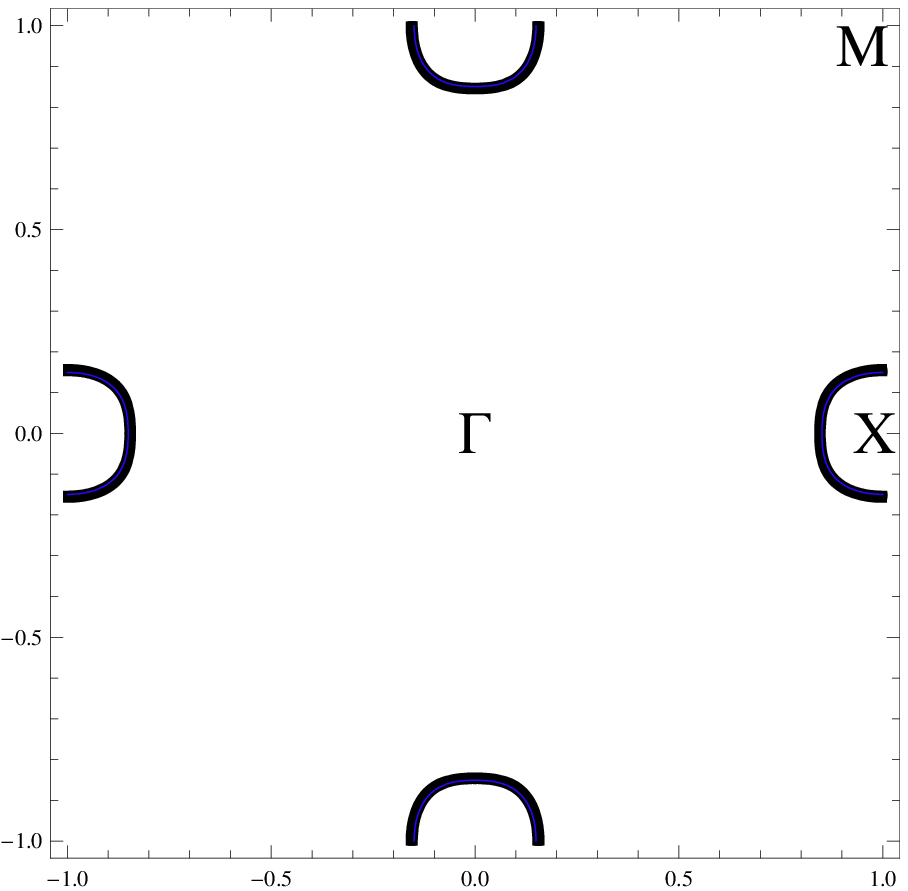}
\caption{Left: A tight-binding fit to the band dispersion observed in Ref.\cite{xj}. Right panel: the electron-like Fermi surfaces.\label{band}}
\end{figure}

\begin{figure}
\includegraphics[scale=.56]{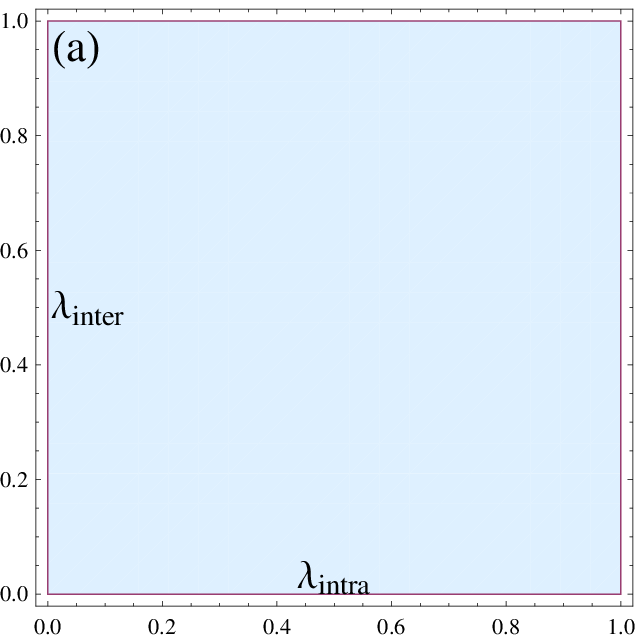}\hspace{0.5in}\includegraphics[scale=.275]{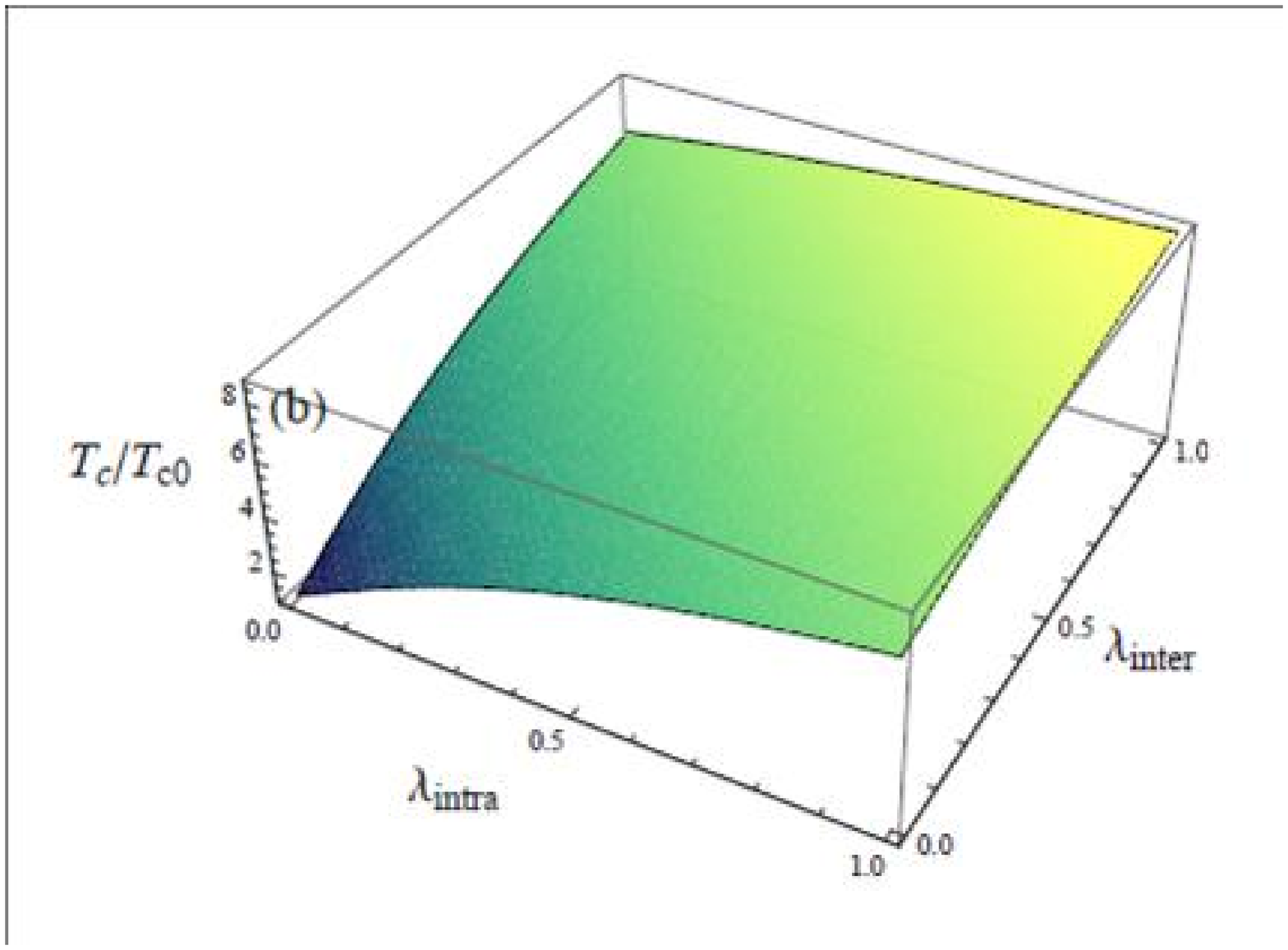}
\caption{(a) the phase diagram. Light blue denotes the  even sign
pairing. (b) the $T_c$ enhancement factor $T_c/T_{c0}$ where
$T_{c0}$ is the superconducting transition temperature in the
absence of the electron-phonon interaction. The parameters we used
to construct the figures are $\Lambda_e/\Lambda_{ph}=2$,
$\langle\w\rangle/\Lambda_{\rm ph}=0.5$, and $V_{\rm intra}=-0.2$,
$V_{\rm inter}=-0.05$.\label{eleph2}}
\end{figure}

Our tight-binding fit of the ARPES bandstructure and the associated Fermi surfaces are  shown in Fig.\ref{band}.
Because the hole bands have completely sunk below the Fermi
energy, the antiferromagnetic fluctuation only occurs at energies
greater than the separation between the top of the hole bands and
the Fermi energy. According Ref.\cite{xj} this separation is
approximately 80 meV. If sufficiently strong this high energy
magnetic fluctuation can trigger $S_{++}$ pairing on the electron
pockets. This pairing form factor can be thought of as $S_{+-}$
restricted to the exposed Fermi surfaces. When this form factor is
the leading pairing channel, we expect $V_{\rm intra}$ and $V_{\rm
inter}$ to be both negative. Under this condition adding the
electron-phonon interaction the phase diagram is shown in
Fig.~\ref{eleph2}(a)- there is only $S_{++}$ phase. The $T_c$
enhancement is shown in Fig.~\ref{eleph2}(b).

For larger separation between the hole bands and $E_F$, the nodeless d-wave (where the gap function
has opposite sign on the two electron pockets) is the leading pairing channel. As pointed out in Ref.\cite{Mazin2}
in the presence of  hybridization between the electron pockets (due to the absence of the $z\leftrightarrow -z$ symmetry at the interface)
the d-wave pairing can become nodal. For strong hybridization the reconstructed electron pockets could have opposite sign pairing due
to the repulsive electron-electron interaction.\cite{Mazin2,chubukov2} We expect in both cases Fig.~\ref{eleph} should apply, i.e.,
strong enough inter-pocket scattering (the first main effect of the FE phonon in earlier discussions), can destabilize the odd sign pairing
and turn it into even sign. In addition to the effect of phonon the inevitable disorder scattering also tends to destabilize the odd sign
pairing in favor of even sign. 
Thus we strongly believe {\it the electron doped FeSe/STO  have even sign (or $S_{++}$) pairing.}


Now we present the more technical part of the paper.

\section{The first stage - FRG calculation}\label{sec:FRG}

The electronic Hamiltonian consists of the bandstructure part,
$H_{\rm band}$, and the Hubbard-Hundes interaction $H_I$. The
details of $H_I$ can be found in Appendix \ref{app:frg}. For the
undoped FeSe/STO because of the well nested Fermi
surfaces\cite{lu} particle-hole scattering is expected to grow
continuously at low energies. As the result the assumption made
earlier, that we only need to retain the Cooper channel
electron-electron scattering for $E<\Lambda_{\rm ph}$, is not a
priori justified. Consequently in the next subsection we shall
perform a full FRG treatment of both electrons and FE phonons to
check the results obtained using the Eliashberg approach.

\subsection{The undoped FeSe/STO}
\begin{figure}
\includegraphics[width=10cm]{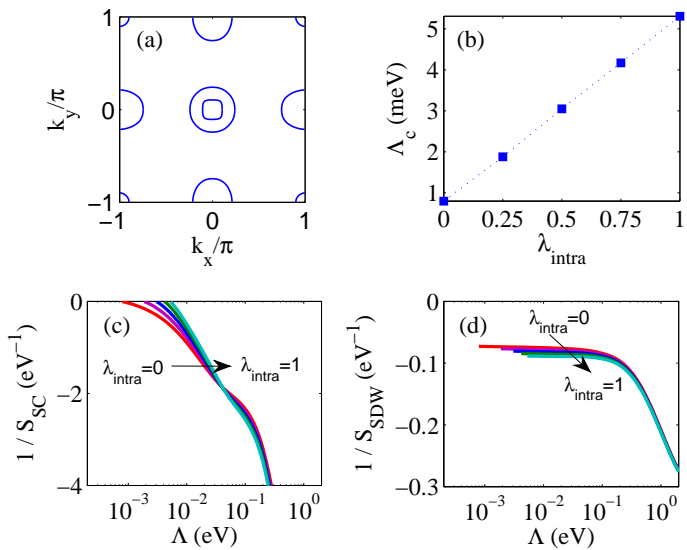}
\caption{(a) FeSe Fermi surface(s). (b) The superconducting
critical scale $\La_c$ versus $\la_{{\rm intra}}$. (c) and (d)
presents the flow of $S_{SC}$ and $S_{SDW}$, respectively. (Higher
scales are not shown). The value of $\la_{{\rm intra}}$ increases
by equal intervals along the arrows. \label{frg-phonon}}
\end{figure}

The Fermi surfaces (in the unfolded Brillouin zone) of the undoped
system are shown in Fig.~\ref{frg-phonon}(a). The phonon
Hamiltonian is given by $H_{\rm ph}=\sum_{\v p} \w(\v p) a_{\v
p}^\dag a_{\v p}$. Here $\v p$ is the three-dimensional momentum.
We assume the following dispersion for the FE phonon $\w(\v
p)=\sqrt{\w_0^2+c^2 p^2}$ where $\w_0\approx 2\mathrm{meV}$  and
$c\sim 70\mathrm{meV}\cdot\AA$ are estimated from an early neutron
measurement.\cite{neutron-STO} The electron-phonon Hamiltonian is
given by \eqa H_{{\rm e-ph}}=A\sum_{i,\al} u_i n_{i\al}\ra
\sum_{p_z}\sum_{\v ka,\v k'b} g_{\v ka,\v k'b; p_z} (a_{\v k-\v
k'+\v p_z}^\dag + a_{\v k'-\v k-\v p_z}) \Psi_{\v k' b}^\dag
\tau_3\Psi_{\v k a}. \label{heph}\eea $H_{\rm e-ph}$ describes the
coupling between electrons in the FeSe layer and the nearest ion
displacement $u_i$ in the TiO$_2$ layer of STO. In the above $\v
p_z=p_z\hat{z}$ is the out-of-plane momentum of the phonon and
$\tau_3$ is the third Pauli matrix in the Nambu space. The
summation over $p_z$ follows from the fact that the
elelctron-phonon coupling occurs at the interface. 
Here $a$ and $b$ label the electron bands, and $g_{\v ka,\v k'b;
p_z} \propto A\<\v k a|\v k'b\>/\sqrt{\w(\v k-\v k'+\v p_z)}$
where $\<\v k a|\v k' b\>$ is the overlap between band Bloch
states. The phonon mediated intra and inter-pocket Cooper
scattering strength are given by $\la_{\rm
intra}=\sum_{a=1}^5\la_{aa}/5$ and $\la_{\rm inter}=\sum_{a\neq
b}\la_{ab}/20$, where \eqa \la_{ab}=\sqrt{N_a
N_b}\sum_{p_z}\Huge{\langle}\Huge{\langle}\frac{2|g_{\v ka,\v k'b;
p_z}|^2}{\w(\v k-\v k'+\v
p_z)}\Huge{\rangle}\Huge{\rangle}_{a,b},\eea where $\<\<\cdot\>\>$
denotes the joint
average over $\v k$ and $\v k'$ which lie on Fermi pockets $a$ and $b$ respectively. 
Using $\w_0$ and $c$ above and the Fermi pockets in
Fig.~\ref{frg-phonon}(a) we estimate $\la_{\rm inter}/\la_{\rm intra}\sim 1/12$. 

We generalize the SM-FRG method\cite{husemann,wang1} to
treat the effects of both electron-electron and electron phonon interactions.
In principle, we envision a boson-fermion FRG calculation that
involves the flow of electron self energy, the phonon self energy,
the elelctron-phonon coupling vertex, the 4-point phonon vertex, as well as
the electron-electron interaction vertex. However in the following shall view the electron dispersion, the phonon
dispersion, and the elelctron-phonon coupling as fully renormalized
quantities. The first two can be determined from experiments and the last quantity is viewed as an adjustable parameter in our theory.  
What's left is the RG flow of the electron-electron
interaction  including contributions from the pure electron-electron interaction and the phonon mediated interaction. In the
following the  renormalized interactions in the
superconducting (SC) and spin density wave (SDW) channels are
denoted as $S_{SC,SDW}$. (The charge density wave channel turns
out to be unimportant and will not be discussed.) The
definitions as well as technical details of the SM-FRG
method can be found in Appendix \ref{app:frg}.

Fig.~\ref{frg-phonon}(b) summarizes the superconducting critical
scale $\La_{c}$ (filled squares) versus the electron-phonon
interaction parameter $\la_{\rm intra}$ ($\lambda_{\rm inter}\sim
\lambda_{\rm inter}/12$). This is extracted from
Fig.~\ref{frg-phonon}(c) which shows the flow of $S_{SC}$.
Fig.~\ref{frg-phonon}(b) shows a linearly rising $\Lambda_c$ as a
function of $\la_{\rm intra}$. When we compare the maximum
$\Lambda_c$ with that in the absence of electron-phonon
interaction a maximum $T_c$ enhancement $\sim$ 6.5 is obtained.
Given the fact that $\lambda_{\rm intra}\sim 12 \lambda_{\rm
inter}$ this result is consistent with Fig.~\ref{eleph}. We have
checked that for all values of $\la_{\rm intra}$ in
Fig.~\ref{frg-phonon}(c) the pairing symmetry remains to be
$S_{+-}$. Fig.~\ref{frg-phonon}(d) shows that the $S_{SDW}$ almost
saturates at low energy scales, and the electron-phonon coupling
reduces $1/S_{SDW}$ only slightly. Since this type of SDW is
related to the pairing interaction, the above observation
justifies our previous assumptions for the input to the Eliashberg
calculation.

\subsection{The electron doped FeSe/STO}

In this case the assumption of only retaining the Cooper channel
electron-electron scattering is a good approximation. Hence we
shall focus on the pure electronic calculation to find out the
leading pairing channels and use it as input for the second
stage Eliashberg calculation. The electronic structure we use to
describe the electron doped FeSe is shown in Fig.~\ref{band}. It
it worthy to note that this bandstructure differs substantially
from those for other pnictides in that the hole band top near
$\Gamma$ and the electron band bottom near X and Y are separated
by a small gap. Interesting in a recent ARPES work on
$A_xFe_ySe_2$\cite{donglai} this feature is noted and emphasized.

\begin{figure}
\includegraphics[scale=0.5]{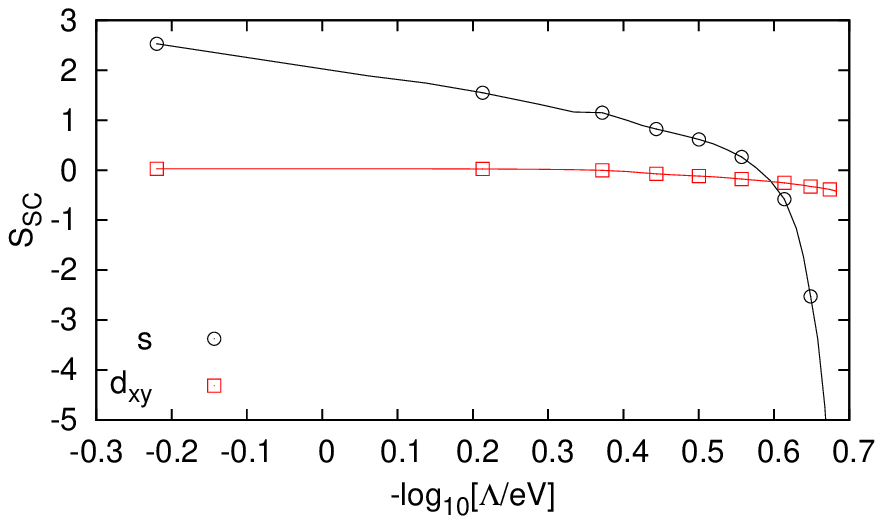}
\includegraphics[scale=0.5]{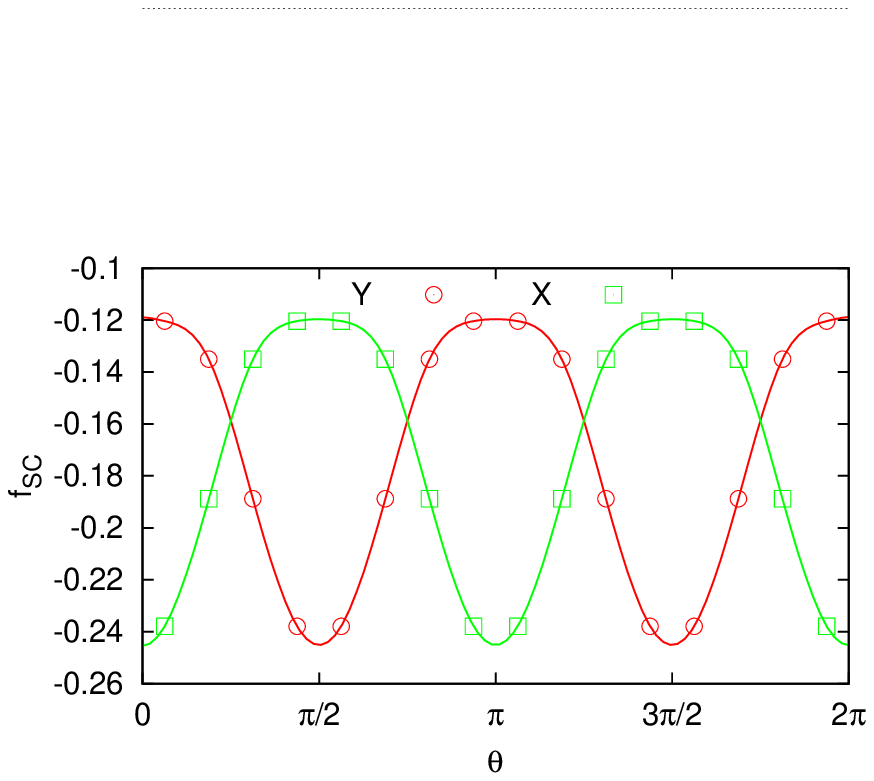}\includegraphics[scale=0.5]{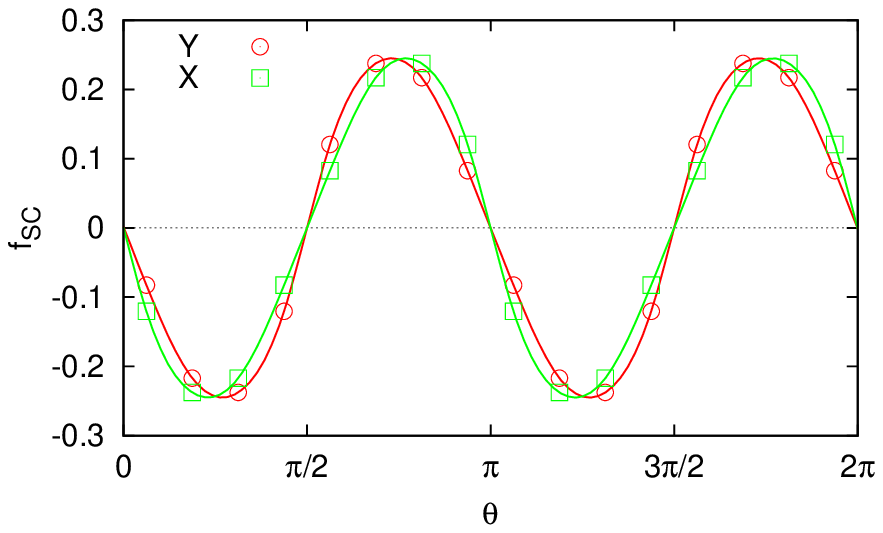}
\caption{Left panel: the RG flow of the Cooper scattering
strength $S_{SC}$
associated with the two leading pairing channels.
Black circles: s-wave pairing. Red squares: $d_{xy}$ pairing.
Mid panel: the s-wave form factor.
The polar angle $\theta$ is defined with respect to the $+x$ direction for both
$X(\pi,0)$ and $Y(0,\pi)$ electron pockets.
Right panel: the $d_{xy}$ form factor.
\label{edope}
}
\end{figure}

For $(U,U',J_H)=(2,1.7,0.15)$ eV, the FRG results for the Cooper pairing is shown in Fig.~\ref{edope}.
The left panel shows the leading pairing instability being in the
$s$-wave channel (black), while the subleading pairing channel has
$d_{xy}$ symmetry (red). The form factors for the $s$ and $d_{xy}$
pairing are shown in mid and right panels respectively. For the
$s$-wave pairing, the  form factor has the same sign on the two
electron pockets X and Y. However the amplitude of the form factor
is quite anisotropic. In contrast the $d_{xy}$ form factor has
nodes on the two electron pockets. We have also checked for
considerably weaker interaction parameters, the leading pairing
channel is $d_{x^2-y^2}$ where the gap function has opposite sign
on the X and Y electron pockets. As discussed earlier, the
interaction with FE phonon tends to stabilize the $s$ wave pairing
and suppress the $d$-wave pairing. For intermediate
electron-electron interaction parameters we obtain $s$-wave
pairing. The addition of FE phonons will further enhance the
pairing energy scale.

\section{The second stage - the Eliahberg equations}

The formalism and notations in this section closely follow those in
Scalapino {\it et al.}\cite{Scalapino66} and
McMillan~\cite{McMillan68}. The details can be found in Appendix \ref{app:Eliashberg}.
In order to make analytic calculation
feasible, we model the low energy electronic degrees of freedom by
two Fermi surfaces. In the case of undoped FeSe these correspond to hole and electron pockets and for  the electron doped case both Fermi surfaces correspond to electron pockets.
 These two Fermi surfaces are labeled by $a=1,2$ in
(\ref{equ:H}) respectively. Moreover we shall assume these Fermi
surfaces have constant density of states(DOS) $N_{1,2}$. In
addition, in the rest of this section the interaction
$V_{abcd}(\boldsymbol{p}_1,\boldsymbol{p}_2,\boldsymbol{p}_3,\boldsymbol{p}_4)$
will be assumed to depend on the Fermi surface index only, {\it
i.e.}, $V_{abcd}$. These simplifications  are made in view of the
presumably strong disorder-induced
quasiparticle scattering at the FeSe/STO interface which will average out the fine structures in
$V_{abcd}(\boldsymbol{p}_1,\boldsymbol{p}_2,
\boldsymbol{p}_3,\boldsymbol{p}_4)$ and the DOS.

Assume the following form for the self energy of the Nambu spinor
$\Psi$,
\begin{equation}
\Sigma_a(\omega)=[1-Z_a(\omega)]\omega\tau_0+Z_a(\omega)\Delta_a(\omega)\tau_1,
\label{equ:Sigma}
\end{equation}
we derive the self-consistent equations for $Z$ and $\Delta$
following standard procedures.\cite{Scalapino66} The results are
given by Eqs.~(\ref{equ:SCEZ}) and (\ref{equ:SCEDelta}) in
Appendix~\ref{app:Eliashberg}. Assuming the McMillan ansatz (Eqs~(\ref{equ:Mansatz})) these equation can be solved to yield $Z_a(0)=1+\sum_{b} \sqrt{N_b/N_a}\lambda_{a b}$
for the normal state $Z_a(0)$,
and the following eigenvalue problem for $T_c$(\ref{equ:SCEDelta})
\begin{equation}
\begin{split}
&
\sum_{b=1}^{2}
\lambda_{ab}
\left [
\ln\frac{\Lambda_{\rm ph}}{T_c}\cdot \sqrt{N_b}\Delta_b(0)
+\frac{\langle \omega\rangle_{ab}}{\Lambda_{\rm ph}}
\cdot \sqrt{N_b}\Delta_b(\infty)
\right ]
+\sqrt{N_a}\Delta_a(\infty)
=\sqrt{N_a}Z_a(0)\Delta_a(0),
\\ &
-\sum_{b=1}^{2}
v_{ab}
\left [
\ln\frac{\Lambda_{\rm ph}}{T_c}\cdot \sqrt{N_b}\Delta_b(0)
+
\ln\frac{\Lambda_{e}}{\Lambda_{\rm ph}}\cdot \sqrt{N_b}\Delta_b(\infty)
\right ]
=\sqrt{N_a}\Delta_a(\infty).
\end{split}
\label{equ:McMillan}
\end{equation}
In the above equations $\lambda_{a b}=2\int_{0}^{\infty}\frac{\mathrm{d}\nu}{\nu}$ and $\alpha_{a b}^2(\nu)F(\nu)$, where $F(\nu)$ is the density of states associated with the
 phonon mode and $\alpha_{a b}^2(\nu)$ is the effective
electron-phonon coupling constants. In addition $v_{ab}=\sqrt{N_a N_b}V_{abba}$, and $\langle\omega\rangle_{ab}$
is a weighted average [weighted by $\alpha_{
ab}^2(\nu)F(\nu)/\nu$] of the phonon
frequency. See
Appendix~\ref{app:Eliashberg} for details.

For the sake of simplicity we shall set $N_1=N_2$,
$\lambda_{aa}\ra\lambda_{\rm intra}=(\lambda_{11}+\lambda_{22})/2$,
$\lambda_{ab}\ra \lambda_{\rm inter}=(\lambda_{12}+\lambda_{21})/2$,
$v_{aa}\ra v_{\rm intra}=(v_{11}+v_{22})/2$, $v_{ab}\ra v_{\rm inter}=(v_{12}+v_{21})/2$, and
$\langle\omega\rangle_{ab}=\langle\omega\rangle$ in the following.
With the above simplifications we can determine the $T_c$ for
$S_{+-}$ ($\Delta_1=-\Delta_2$) and $S_{++}$ ($\Delta_1=\Delta_2$)
pairing as given by Eq.~(\ref{equ:tc}).  We have checked changing $N_1/N_2$ has little effect on $T_c$ while
have a strong effect on $\Delta_1/\Delta_2$.

If phonons are indeed responsible for the observed $T_c$
enhancement, there should be other signatures of the
electron-phonon coupling. For example due to the large anharmonicity of the FE phonons we expect
using a pump laser to excite them will have a clear effect on $T_c$. The more traditional phonon signatures
such as kink in the quasiparticle dispersion and shoulder in the tunneling experiment are discussed in
Appendix~\ref{app:kink}.

\section{Discussion}

We have studied the screening effects the ferroelectric phonons of
SrTiO$_3$ on the interaction between the electrons in FeSe. We
conclude such coupling can enhance the pairing strength of FeSe.
Moreover we find when the inter-pocket electron-phonon scattering
is strong, the opposite sign pairing will give way to the equal
sign pairing.

An immediate question one might ask is why don't the FE phonons
have similar effect on the $T_c$ of doped STO. We believe the
answer is polaron formation - for a range of strong electron-phonon
coupling, the formation of polarons instead of Cooper pairs are
favored. In a recent optical experiment on n-type doped
STO,\cite{pol} a very sharp Drude peak with a substantial mass
enhancement (consistent with that of ``large polarons'') was
observed.

The current study raises the concern about whether the role of
phonon can be completely ignored in bulk iron-based superconductors.\cite{kon} 
With appropriate interpretation of $\lambda_{\rm intra}$ and
$\lambda_{\rm inter}$, our results can be used to
address the phonon effects in bulk iron-based superconductors as
well. 

Material wise there
are other nearly ferroelectric perovskite materials, for example
KTaO$_3$.\cite{KTaO3} If FeSe films can be epitaxially grown on
these materials similar $T_c$ enhancement should occur. Finally the
results of Ref.\cite{xue} and the present paper suggest the
... FeSe/(STO)$_n$/FeSe/(STO)$_n$... superlattice is a promising
artificial material with high $T_c$.

\acknowledgments{ We are in debt to Yuan-Ming Lu who helped us
understand the phonons in SrTiO$_3$, and Fan Yang for informing us
of the experimental result of Ref.\cite{xue}. We thank Qi-Kun Xue
and Xingjiang Zhou for sharing their unpublished results with us,
and  R. Ramesh for
telling us many important properties of SrTiO$_3$. We also thank
Todadri Senthil, Tao Xiang, Jun Zhao and Yuan Wan for helpful discussions.
QHW acknowledges the support by the Ministry of Science and
Technology of China (under grant No.2011CBA00108 and 2011CB922101)
and NSFC (under grant No.10974086, No.10734120 and No.11023002).
DHL acknowledges the support by the DOE grant number
DE-AC02-05CH11231.}

\begin{appendix}

\section{The SM-FRG method}
\label{app:frg}

In this section we describe the SM-FRG method we used. The
microscopic hamiltonian, which is valid for all cutoff $<$
bandwidth, we use is $H=H_{\rm band}+H_I+H_{\rm ph}+H_{{\rm
e-ph}}$.  Here $H_{\rm band}$ is the two dimensional
bandstructure. For the undoped case $H_{\rm band}$ is a courtesy
of Z-Y Lu.\cite{zy} For the electron doped case it obtained by a
fit to the ARPES result\cite{xj}.  $H_I$ describes the local
electron-electron interaction given by \eqa H_I=&&U\sum_{i,\al}
n_{i,\al,\ua}n_{i,\al,\da}+U'\sum_{i,\al>\bt}n_{i,\al} n_{i,\bt} +
J_H
\sum_{i,\al>\bt,\si,\si'}\psi_{i,\al,\si}^\dag\psi_{i,\bt,\si}\psi_{i,\bt,\si'}^\dag\psi_{i,\al,\si'}
\nn &&+ J_H
\sum_{i,\al>\bt}(\psi_{i,\al,\ua}^\dag\psi_{i,\al,\da}^\dag\psi_{i,\bt,\da}\psi_{i,\bt,\ua}+{\rm
h.c.}),\eea where $\psi_{i,\al,\si}$ annihilates a spin $\si$
electron at site $i$ in orbital $\al$
($\al=3z^2-r^2,xz,yz,x^2-y^2,xy$),
$n_{i,\al,\si}=\psi_{i,\al,\si}^\dag\psi_{i,\al,\si}$ and
$n_{i,\al}=\sum_\si n_{i,\al,\si}$. In the calculation of the
main text we used intra-orbital repulsion $U=2eV$, Hund's rule
coupling $J_H=0.31eV$, and inter-orbit repulsion $U'=U-2J_H$.

\begin{figure}
\includegraphics[width=8cm]{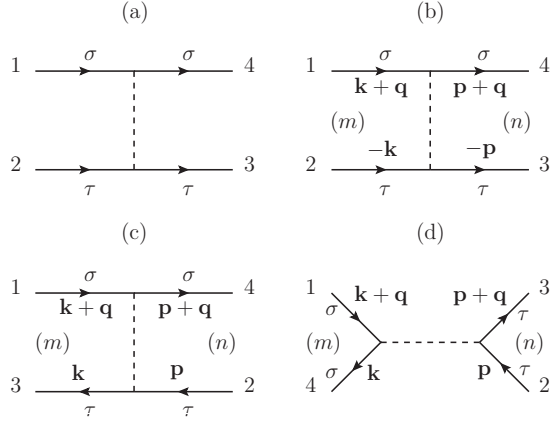}
\caption{A generic 4-point vertex (a) is rearranged into the
pairing (b), crossing (c) and direct (d) channels. Here $\v k,\v
q,\v p$ are momenta, $\si$ and $\tau$ denote spins which are
conserved during fermion propagation, and $m,n$ denote the form
factor (see the text for details).}\label{pcd}
\end{figure}

Fig.~\ref{pcd} (a) shows a generic 4-point vertex function
$\Ga_{1234}$ which appears in the interaction $\psi^\dagger_1
\psi^\dagger_2 (-\Ga_{1234}) \psi_3 \psi_4$. Here $1,2,3,4$
represent momentum (or real space position) and orbital label. The
spin $\si$ and $\tau$ are conserved along fermion propagators and
will be suppressed henceforth. Figs.\ref{pcd}(b)-(d) are
rearrangements of (a) into the pairing (P), the crossing (C) and
the direct (D) channels in such a way that a collective momentum
$\v q$ can be identified. The dependence on all other momenta and
orbital labels is written as \eqa && \Ga^{\al\bt\ga\del}_{\v k+\v
q,-\v k,-\v p,\v p+\v q}\ra \sum_{mn}f_m^*(\v k,\al,\bt)P_{mn}(\v
q)f_n(\v p,\del,\ga),\nn && \Ga^{\al\bt\ga\del}_{\v k+\v q,\v p,\v
k,\v p+\v q}\ra \sum_{mn}f_m^*(\v k,\al,\ga)C_{mn}(\v q)f_n(\v
p,\del,\bt),\nn&& \Ga^{\al\bt\ga\del}_{\v k+\v q,\v p,\v p+\v q,\v
k}\ra \sum_{mn}f_m^*(\v k,\al,\del)D_{mn}(\v q)f_n(\v
p,\ga,\bt).\label{projection} \eea Here $f_{m=(l,o)}(\v
k,\al,\bt)=h_l(\v k){\cal{M}}_o(\al,\bt)$ is a composite form
factor, where $h_l(\v k)$ is chosen from a set of orthonormal
lattice harmonics, ${\cal{M}}_o$ is a matrix in the orbital basis.

The decomposition Eq.~(\ref{projection}) for each channel would be
exact if the form factor set is complete. In practice, however, a
set of a few form factors is often sufficient to capture the
symmetry of the order parameters associated with leading
instabilities.\cite{husemann,wang1} In our case, the lattice
harmonics are chosen as $h(\v k)=1$, $\cos k_x\pm \cos k_y$,
$2\cos k_x\cos k_y$ and $2\sin k_x\sin k_y$. They are all even
since only singlet pairing is relevant in our case. The
${\cal{M}}$-matrices are chosen so that the combination
$\sum_{\al\bt}\phi_\al{\cal{M}}(\al,\bt)\phi_\bt$ ($\phi_\al$ is
the real atomic orbital function) is irreducible and transforms
according to $A_{1g}$, $B_{1g}$ or $B_{2g}$ under the point
group.\cite{wan} (One may also use any bilinear $\phi_\al\phi_\bt$
to determine a matrix ${\cal M}$, but it is less transparent
symmetry wise.) Moreover, the ${\cal{M}}$-matrix is normalized as
$\Tr {\cal{M}}^\dag {\cal{M}}=1$.  If the total number of
composite form factors is $N$, then $P$, $C$ and $D$ are all
$N\times N$ matrix functions of $\v q$. Note that the $P,C$ and
$D$ channels are not orthogonal. The overlap between different
channels are important for the growth of pairing interaction out
of, e.g., the magnetic interaction.\cite{wl,wang1,frg} In the
following we denote $X_K=\hat{K}X$  as the projection of $X$ into
the K-channel via Eq.~(\ref{projection}).

\begin{figure}
\includegraphics[width=8cm]{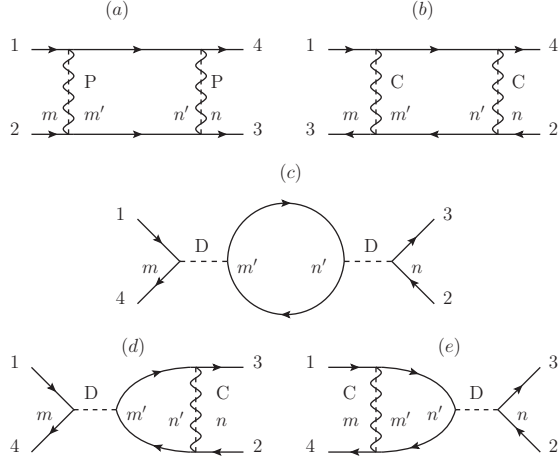}
\caption{One-loop diagrams contributing to the flow of the the
4-point vertex function in the pairing channel (a), crossing
channel (b), and direct channel (c)-(e). Here $m,m'n,n'$ denote
form factors, while the momentum, orbital and spin indices are
left implicit. The dashed line represent 4-point fermion vertex,
dash-wavy line means that both fermion vertex and phonon
propagator can be inserted but separately. The diagrams are
one-particle-irreducible with respect to both fermions and
phonons, and the phonon line shares the loop frequency since
external fermion fields are set at zero frequency as usual. The
Matsubara frequency is continuous but subject to hard infrared
cutoff at running scale $\La$. }\label{oneloop}
\end{figure}

In the case with electron-phonon interaction we need the phonon-mediated electron-electron scattering.
According to the $H_{\rm e-ph}$
defined in the text, this vertex is given by, \eqa V(\v q
,\nu_n)\propto \sum_{p_z} \frac{A^2}{\nu_n^2+\w^2(\v
q+p_z\hat{z})}\propto \frac{A^2}{\sqrt{\nu_n^2+\w^2(\v q)}},\eea
where $\v q=(q_x,q_y)$ and $\nu_n$ are the momentum and
(Matsubara) frequency transfer in the electron-electron
scattering, and the last proportionality holds to leading order in
$\w_0/c$. The $\la_{\rm intra,inter}$ discussed in the main text are both proportional to $A^2$. 
Notice that this vertex is naturally in the $D$-channel and bears
trivial form factors with $h=1$ and ${\cal M}=\del_{\al\bt}$ since
according to Eq.~(\ref{heph}) the electron-phonon interaction is
local in real space and diagonal in orbital basis.

The partial flows of $P$, $C$ and $D$ are given by the
one-particle-irreducible diagrams shown in Fig.~\ref{oneloop}. Here
the dashed line denotes 4-point fermion vertex, wavy line the
(surface) phonon-mediated vertex, and the dash-wavy line means
that both types of vertices can enter. We write the partial flow
equations as, in matrix form and for a collective momentum $\v q$,
\eqa && \frac{\p P}{\p\La} = (P+V_P)\chi'_{pp}(P+V_P),\nn
&&\frac{\p C}{\p\La} =(C+V_C)\chi'_{ph}(C+V_C),\nn && \frac{\p
D}{\p\La}=(C+V_C-D)\chi'_{ph}D+D\chi'_{ph}(C+V_C-D).
\label{pflow}\eea Here $\La$ is the running Matsubara frequency
cutoff, $V_K$ is the phonon mediated interaction projected
into the $K$ channel at cutoff $\La$, and $\chi'_{pp/ph}$ are
matrix-kernels with elements
\begin{widetext} \eqa
(\chi'_{pp})_{mn}&&=-\frac{1}{2\pi}\int\frac{d^2\v
p}{(2\pi)^2}f_m(\v p,\al,\bt)G_{\al\ga}(\v p+\v
q,i\La)G_{\bt\del}(-\v p,-i\La)f_n^*(\v p,\ga,\del)\ \ +(\La\ra
-\La),\nn (\chi'_{ph})_{mn} &&=-\frac{1}{2\pi}\int\frac{d^2\v
p}{(2\pi)^2}f_m(\v p,\al,\bt)G_{\al\ga}(\v p+\v
q,i\La)G_{\del\bt}(\v p,i\La)f_n^*(\v p,\ga,\del)\ \ +(\La\ra
-\La),\label{loopint} \eea
\end{widetext} where $G$ is the bare fermion propagator in the
obital basis, and the summation over orbitals is left implicit.

Clearly, because of the $\La$ dependence, the effect of phonon
mediated interaction, $V$, is important only if $\La$ reaches the
phonon band, above which the main contribution to the flow of the
fermion interaction vertex is from pure electron-electron
interaction. However the electronic excitations do modify phonon
self energy and elelctron-phonon vertex even when the cutoff scale
is above the phonon bandwidth. In the present work such effects
are accounted for by using the experimentally measured phonon
dispersion and the elelctron-phonon coupling constant. The flows in
Eq.~(\ref{pflow}) collect contributions from independent
one-particle-irreducible diagrams for the total change $d\Ga$,
which need to be subsequently projected to the three channels.
Therefore the full flow equations can be formally written as, \eqa
\frac{dK}{d\La}=\frac{\p K}{\p\La} + \hat{K}\sum_{K'\neq
K}\frac{\p K'}{\p\La},\label{fullflow}\eea for $K=P,C$ and $D$. We
used the fact that $\hat{K}\p K=\p K$ by definition.

The functions  $P$, $C$ and $D$ are related to the effective
interactions as, $V_{sc}=-P-V_P$ in SC-, $V_{sdw}=C+V_C$ in SDW-,
and $V_{cdw}=C+V_C-2D$ in CDW-channels. We monitor the most
negative singular values $S_{SC,SDW,CDW}$ of such interactions
(for all $\v q$) versus the running cutoff $\La$. The most
negative one among $S_{SC,SDW,CDW}$ tells us which channel is
becoming unstable. The associated eigen function dictates the
symmetry and wave vector of the order parameter.

\section{Multiple band Eliashberg equation}
\label{app:Eliashberg} In this Appendix we briefly outline the
derivation of multiple band Eliashberg eqautions for the
electron-phonon Hamiltonian (\ref{equ:H}), following the single
band case of Scalapino {\it et al.}~\cite{Scalapino66}

The effective elelctron-phonon Hamiltonian we consider, at the
energy cutoff $\Lambda_e$, is given by
\begin{equation}
\begin{split}
H=
\ &
\sum_{a=1}^{2}\sum_{\boldsymbol{p}}
\epsilon_{\boldsymbol{p}a}
\Psi_{\boldsymbol{p}a}^\dagger \tau_3\Psi_{\boldsymbol{p}a}^{\vphantom{\dagger}}
+\sum_{\boldsymbol{q}}
\w(\boldsymbol{q})
a_{\boldsymbol{q}}^\dagger a_{\boldsymbol{q}}^{\vphantom{\dagger}}
\\ &
+\sum_{a,b}\sum_{\boldsymbol{p},\boldsymbol{p}'}
g_{\boldsymbol{p}\boldsymbol{p}',a b}
\varphi_{\boldsymbol{p}-\boldsymbol{p}'}
\Psi_{\boldsymbol{p}'b}^\dagger \tau_3\Psi_{\boldsymbol{p}a}^{\vphantom{\dagger}}
\\ &
+\frac{1}{2}\sum_{a,b,c,d}
\sum_{\boldsymbol{p}_1,\boldsymbol{p}_2,\boldsymbol{p}_3}
V_{abcd}(\boldsymbol{p}_1,\boldsymbol{p}_2,\boldsymbol{p}_3,
\boldsymbol{p}_4)
(\Psi_{\boldsymbol{p}_3 c}^\dagger \tau_3\Psi_{\boldsymbol{p}_1 a}^{\vphantom{\dagger}})
(\Psi_{\boldsymbol{p}_4 d}^\dagger \tau_3\Psi_{\boldsymbol{p}_2 b}^{\vphantom{\dagger}}).
\end{split}
\label{equ:H}
\end{equation}
Here $\Lambda_e$ is much smaller than the bandwidth but
larger than maximal phonon frequency $\Lambda_{\rm ph}$,
$\Psi_{\boldsymbol{p} a}$ is the Nambu spinor for electron band
$a$ with dispersion $\epsilon_{\boldsymbol{p}a}$,
$a_{\boldsymbol{q}}$ is the destruction operator for phonon
with dispersion $\w(\boldsymbol{q})$,
$\varphi_{\boldsymbol{q}}= a_{\boldsymbol{q}}^\dagger +
a_{-\boldsymbol{q}}^{\vphantom{\dagger}}$,
$\boldsymbol{p}_4=\boldsymbol{p}_1+\boldsymbol{p}_2-\boldsymbol{p}_3$,
$\tau_3$ is the Pauli matrix, and
$V_{abcd}(\boldsymbol{p}_1,\boldsymbol{p}_2,\boldsymbol{p}_3,\boldsymbol{p}_4)$
is the (effective) electron-electron interaction. Different levels
of simplification will be applied to  Eq.~(\ref{equ:H}), which
will be discussed in more details in the following.

Assume the
Green's function of Nambu spinor $\Psi$ to be given by
\begin{equation}
[G_a(\boldsymbol{p},\omega)]^{-1}
=\omega-\epsilon_{\boldsymbol{p} a}\tau_3-\Sigma_a(\boldsymbol{p},\omega).
\end{equation}
The self-consistent equation of
the self-energy $\Sigma_a(\boldsymbol{p},\omega)$ is
\begin{equation}
\begin{split}
\Sigma_a(\boldsymbol{p},i \omega)
=
-T\sum_{\omega',\boldsymbol{p}'}
\sum_{b}
\tau_3 G_b(\boldsymbol{p}',i \omega') \tau_3\cdot
\Big [
&
|g_{\boldsymbol{p}\boldsymbol{p}',a b}|^2\,
D(\boldsymbol{p}-\boldsymbol{p}',i \omega-i \omega')
\\&
+
V_{abba}(\boldsymbol{p},\boldsymbol{p}',\boldsymbol{p}',\boldsymbol{p})
\Big ]
\end{split}
\ee
where $\omega'$ is fermion Matsubara frequency.
$D(\boldsymbol{q},\nu)$ is the Green's function of the  phonon.
Use the spectral representation
\begin{equation}
D(\boldsymbol{q},i \nu')=
\int_0^{\infty}
\mathrm{d} \nu\,
B(\boldsymbol{q}, \nu)\left [
\frac{1}{i \nu'-\nu}-\frac{1}{i \nu'+\nu}
\right ],
\end{equation}
and sum over $\omega'$ by the procedure of Ref.~\cite{Scalapino66},
the self-consistent equation becomes
\begin{equation}
\begin{split}
\Sigma_{a}(\boldsymbol{p},\omega)
=
\ &
-\sum_b \frac{1}{\pi}
\sum_{\boldsymbol{p}'}
\int_{-\infty}^{\infty}
\mathrm{d} \omega''\,
\Im [ \tau_3 G_b(\boldsymbol{p}',\omega'')\tau_3 ]
\\ &\quad
\times
|g_{\boldsymbol{p}\boldsymbol{p}',a b}|^2
\int_{0}^{\infty}\mathrm{d} \nu\, B(\boldsymbol{p}-\boldsymbol{p}',\nu)
\Big [
\frac{N(\nu)+f(-\omega'')}{\omega-\omega''-\nu}
+\frac{N(\nu)+f(\omega'')}{\omega-\omega''+\nu}
\Big ]
\\ &
-\sum_b \frac{1}{\pi}
\sum_{\boldsymbol{p}'}
\int_{-\infty}^{+\infty}
\mathrm{d} \omega''\,
\Im [ \tau_3 G_b(\boldsymbol{p}',\omega'')\tau_3 ]
\frac{1}{2}V_{abba}(\boldsymbol{p}-\boldsymbol{p}')\tanh(\beta \omega''/2)
\end{split}
\end{equation}
Assume each electron band $a$ has a circular Fermi surface
with constant DOS $N_a$,
define
$
\alpha_{a b}^2(\nu)F(\nu)
$
as the average of
$
\sqrt{N_a N_b}
|g_{\boldsymbol{p}\boldsymbol{p}',a b}|^2
B(\boldsymbol{p}-\boldsymbol{p}',\nu)
$
over $\boldsymbol{p}$ on Fermi surface $a$
and $\boldsymbol{p}'$ on Fermi surface $b$,
and ignore the momentum dependence of $\Sigma$
close to Fermi surface,
this equation further simplifies to
\begin{equation}
\begin{split}
\sqrt{N_a}\Sigma_a(\omega)
=
\ &
\sum_b \sqrt{N_b}
\int_{-\Lambda_e}^{\Lambda_e}\mathrm{d} \epsilon_{\boldsymbol{p}',b}\,
\Big \{
\\ &\quad
-\frac{1}{\pi}
\int_{-\infty}^{\infty}\mathrm{d} \omega''\,
\int_{0}^{\infty}\mathrm{d} \nu\,
\alpha_{a b}^2(\nu)F(\nu)
\Im [\tau_3 G_b(p',\omega'')\tau_3]
\\ &\quad\quad\times
\Big [
\frac{N(\nu)+f(-\omega'')}{\omega-\omega''-\nu}
+\frac{N(\nu)+f(\omega'')}{\omega-\omega''+\nu}
\Big ]
\\ &\quad
-\frac{1}{\pi}
\int_{-\infty}^{+\infty}
\mathrm{d} \omega''\,
\Im [ \tau_3 G_b(p',\omega'')\tau_3 ]
\frac{1}{2}v_{ab}\tanh(\beta \omega''/2)
\Big \}.
\end{split}
\end{equation}
Here $N(\nu)=1/(e^{\beta\nu}-1)$ and $f(\omega)=1/(e^{\beta\omega}+1)$ are the Bose and Fermi distribution functions,
$v_{ab}=\sqrt{N_a N_b}V_{abba}$.
Assume that $\Sigma_a$ takes the form of (\ref{equ:Sigma}),
the above equation reduces to
\begin{equation}
\begin{split}
\sqrt{N_a}[1-Z_a(\omega)]=
\ &
\sum_{b}\sqrt{N_b}
\int_{0}^{\infty}\mathrm{d} \nu\,
\alpha_{a b}^2(\nu)F(\nu)\,
\int_{0}^{\Lambda_e}\mathrm{d} \omega'\,
\mathrm{Re}\left[\frac{\omega'}{\sqrt{\omega'^2-\Delta_b^2(\omega')}}\right]
\\ &\quad\times 2\,
\left [
\frac{N(\nu)+f(-\omega')}{\omega^2-(\omega'+\nu)^2}
+
\frac{N(\nu)+f(\omega')}{\omega^2-(\omega'-\nu)^2}
\right ],
\end{split}
\label{equ:SCEZ}
\end{equation}
and
\begin{equation}
\begin{split}
\sqrt{N_a}Z_a(\omega)\Delta_a(\omega)=
\ &
-\sum_{b}\sqrt{N_b}
\int_{0}^{\infty}\mathrm{d} \nu\,
\alpha_{a b}^2(\nu)F(\nu)\,
\int_{0}^{\Lambda_e}\mathrm{d} \omega'\,
\mathrm{Re}\left[\frac{\Delta_b(\omega')}{\sqrt{\omega'^2-\Delta_b^2(\omega')}}\right]
\\ &\quad\times 2\,
\left [
\frac{(\omega'+\nu)[N(\nu)+f(-\omega')]}{\omega^2-(\omega'+\nu)^2}
+
\frac{(\omega'-\nu)[N(\nu)+f(\omega')]}{\omega^2-(\omega'-\nu)^2}
\right ]
\\ &
-\sum_b \sqrt{N_b}
\int_{0}^{\Lambda_e}
\mathrm{d} \omega'\,
\mathrm{Re}\left[\frac{\Delta_b(\omega')}{\sqrt{\omega'^2-\Delta_b^2(\omega')}}\right]
v_{ab}\tanh(\beta \omega'/2).
\end{split}
\label{equ:SCEDelta}
\end{equation}

These equations can be solved
numerically for the frequency dependence of $Z$ and $\Delta$. For
a more transparent demonstration of the physics, we adopt the
McMillan approximation~\cite{McMillan68} and look for solutions of
the form
\begin{equation}
\begin{split}
&
Z_a(\omega<\Lambda_{\rm ph})=Z_a(0),\quad
Z_a(\omega>\Lambda_{\rm ph})=1,\quad
\\
&
\Delta_a(\omega<\Lambda_{\rm ph})=\Delta_a(0),\quad
\Delta_a(\omega>\Lambda_{\rm ph})=\Delta_a(\infty).
\end{split}
\label{equ:Mansatz}
\end{equation}
This leads to the generalized McMillan formula (\ref{equ:McMillan})
in main text.

\section{The phonon signatures}
\label{app:kink}

Conventional signatures of the electron-phonon interaction include the phonon
induced kink in the normal state dispersion and the phonon shoulder in the tunneling
spectra. However these features are most pronounced when  $\alpha^2(\nu)
F(\nu)$ have a sharp peak at a characteristic phonon frequency. While this is indeed the case
for Einstein phonons it is not true for the soft phonons under discussion. Here we expect
$\alpha^2(\nu)
F(\nu)$ to have a wide distribution. Therefore the above phonon features may not be very obvious.
For example using the parameters described in the main text a typical renormalized quasiparticle
dispersion in the normal state near the Fermi surface is shown in
Fig.~\ref{fig:Evsk}.
\begin{figure}
\includegraphics{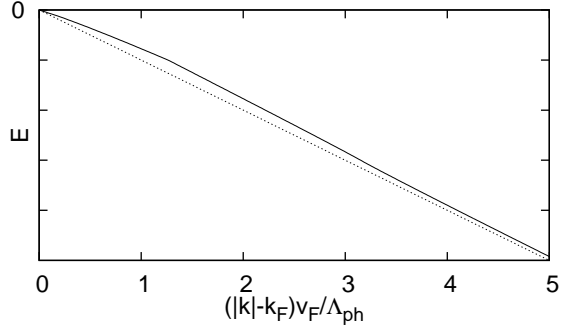}
\caption{Solid(dash) line is the renormalized(un-renormalized)
electron dispersion, obtained by numerical solution of
(\ref{equ:SCEZ}) with parameters $\lambda_{\rm intra}=0.5$ and
$\lambda_{\rm inter}=0$, and the model of FE phonon described in
Section~\ref{sec:FRG}. No prominent kink is visible despite significant
(factor $1.5$) mass enhancement at Fermi level. } \label{fig:Evsk}
\end{figure}

\end{appendix}


\begin{references}
\bibitem{xue} Q.-Y. Wang {\it et al.}, Chin. Phys. Lett. {\bf 29}, 037402 (2012).

\bibitem{xj} Defa Liu {\em{et al}}, arXiv:1202.5849.

\bibitem{lu} Kai Liu, Zhong-Yi Lu, and Tao Xiang, arXiv:1202.0538.

\bibitem{Kivelson}
G. T. Zimanyi, S. A. Kivelson, and A. Luther,
Phys. Rev. Lett. {\bf 60}, 2089 (1988).

\bibitem{Fu} H. Fu, C. Honerkamp and D.-H. Lee, , Europhys. Lett. {\bf 76} 146 (2006).

\bibitem{frg1}C. Honerkamp, M. Salmhofer, N. Furukawa, T. M. Rice,
Phys. Rev. B 63, 035109 (2001).

\bibitem{frg}
F. Wang, H. Zhai, Y. Ran, A. Vishwanath, and D.-H. Lee, Phys. Rev.
Lett. 102, 047005 (2009); F. Wang, H. Zhai, D.-H. Lee, Europhys.
Lett., {\bf 85}, 37005 (2009).

\bibitem{Chubukov}
A. V. Chubukov, D. Efremov, I. Eremin, Phys. Rev. B {\bf 78},
134512 (2008).

\bibitem{wl} For a recent review of theoretical studies of the pairing mechanism
in the iron based superconductors, see  F. Wang and D.-H. Lee,
Science {\bf 332}, 200 (2011).

\bibitem{McMillan68}
W. L. McMillan, Phys. Rev. {\bf 167}, 331 (1968).

\bibitem{Mazin}
I. I. Mazin, D. J. Singh, M. D. Johannes, and M. H. Du, Phys. Rev. Lett. {\bf 101}, 057003 (2008)

\bibitem{Mazin2}
I. I. Mazin, Phys. Rev. B {\bf 84}, 024529 (2011).

\bibitem{chubukov2}
M. Khodas, A. V. Chubukov, arXiv:1202.5563.

\bibitem{neutron-STO}
Y. Yamada, and G. Shirane,
J. Phys. Soc. Jpn. {\bf 26}, 396 (1969).

\bibitem{husemann}
C. Husemann, and M. Salmhofer, Phys. Rev. B {\bf 79}, 195125 (2009).

\bibitem{wang1}
Wan-Sheng Wang, Yuan-Yuan Xiang, Qiang-Hua Wang, Fa Wang,
Fan Yang, and Dung-Hai Lee, Phys. Rev. B {\bf 85}, 035414 (2012).

\bibitem{donglai}
F. Chen, {\it et al.}, Phys. Rev. X {\bf 1}, 021020 (2011).

\bibitem{Scalapino66}
D. J. Scalapino, J. R. Schrieffer, and J. W. Wilkins, Phys. Rev.
{\bf 148}, 263 (1966).







\bibitem{pol}
J. L. M. van Mechelen, D. van der Marel, C. Grimaldi, A. B. Kuzmenko,
N. P. Armitage, N. Reyren, H. Hagemann, and I. I. Mazin,
Phys. Rev. Lett. {\bf 100}, 226403 (2008);
J. L. M. van Mechelen {\it et al.},
Phys. Rev. Lett. {\bf 100}, 226403 (2008).

\bibitem{kon} This issue has been raised in H. Kontani and S. Onari, Phys. Rev. Lett. {\bf 104 }, 157001 (2010).

\bibitem{KTaO3}
C. H. Perry, and T. F. McNelly, Phys. Rev. {\bf 154}, 456 (1967);
C. H. Perry, R. Currat, H. Buhay, R. M. Migoni, W. G. Stirling, and J. D. Axe, Phys. Rev. B {\bf 39}, 8666 (1989).

\bibitem{zy} Zhong-Yi Lu, private communication.

\bibitem{wan} Yuan Wan and Qiang-Hua Wang, EPL {\bf 85}, 57007
(2009). In this paper symmetries of bilinears involving $xz$ and
$yz$ orbitals are considered, but extension to five orbitals is
straightforward.


\end{references}
\end{document}